\documentclass{MolSpaLab}
\usepackage{graphicx}
\begin{document}
\TitreGlobal{Molecules in Space \& Laboratory}
\title{Formation of molecular hydrogen \\
on amorphous silicate surfaces}
\author{
L. Li}
\address{Physics Department, Syracuse University, Syracuse, NY 13244, USA}
\author{
G. Manic\'o}
\address{Universit\'a di Catania, DMFCI, 95125 Catania, Sicily, Italy}
\author{
E. Congiu$^{1,2}$}
\address{Universit\'a di Cagliari, Dipartimento di Fisica, Cagliari Italy}
\author{
J. Roser}
\address{NASA Ames, Mail Stop 245-6, Moffett Field, CA, 94035, USA}
\author{
S. Swords$^1$}
\author{
H.B. Perets}
\address{Faculty of Physics, Weizmann Institue of Science, 
Rehovot 76100, Israel}
\author{
A. Lederhendler}
\address{Racah Institute of Physics, The Hebrew University, 
Jerusalem 91904, Israel}
\author{
O. Biham$^6$}
\author{
J.R. Brucato}
\address{INAF-Osservatorio Astronomico di Capodimonte, Napoli, Italy}
\author{
V. Pirronello$^3$}
\author{
G. Vidali$^1$}

\runningtitle{Formation of molecular hydrogen}

\setcounter{page}{1}

\maketitle

\begin{abstract}

Experimental results on the formation of molecular hydrogen on amorphous 
silicate surfaces are presented and analyzed using a rate equation model. 
The energy barriers for the relevant diffusion and
desorption processes are obtained. 
They turn out to be significantly higher than those obtained 
for polycrystalline silicates,
demonstrating the importance of grain morphology. 
Using these barriers we evaluate the 
efficiency of molecular hydrogen formation on amorphous silicate
grains under interstellar conditions. 
It is found that unlike polycrystalline silicates, amorphous silicate
grains are efficient catalysts of H$_{2}$ formation 
in diffuse interstellar clouds.

\end{abstract}

H$_{2}$ is the most abundant molecule in the interstellar medium (ISM).
It plays a crucial role in the initial cooling of clouds during
gravitational collapse and is involved in most reaction schemes
that produce other molecules.
It is widely accepted that
H$_{2}$ formation in the ISM takes place on the surfaces of dust grains
(Gould \& Salpeter 1963).
In this process,
H atoms that collide with a grain 
quickly equilibrate 
and stick on its surface.
The adsorbed atoms diffuse on the surface.
They may encounter each other and form
H$_{2}$ molecules
(Williams 1968, Hollenbach \& Salpeter 1971, Hollenbach et al. 1971),
or desorb thermally
in atomic form.

In recent years, we have conducted a series of
experiments on molecular hydrogen formation on
dust grain analogues
such as polycrystalline silicates
(Pirronello et al. 1997),
amorphous carbon
(Pirronello et al. 1999)
and amorphous water ice
(Manico et al. 2001, Roser et al. 2002, Perets et al. 2005),
under astrophysically relevant conditions.
In these experiments, the surface was irradiated by beams of H and D atoms.
The production of HD molecules
was measured during the irradiation and during
a subsequent temperature programmed desorption (TPD) experiment.
Related studies were done on
amorphous ice and other surfaces
(Hornekaer et al. 2003, Perets et al. 2005, Dulieu et al. 2005,
Hornekaer et al. 2005, Amiaud et al. 2006, Creighan et al. 2006,
Williams et al. 2007).

\begin{figure}[ht]
\begin{center}
\vspace{0.2in}
\includegraphics[width=8 cm]{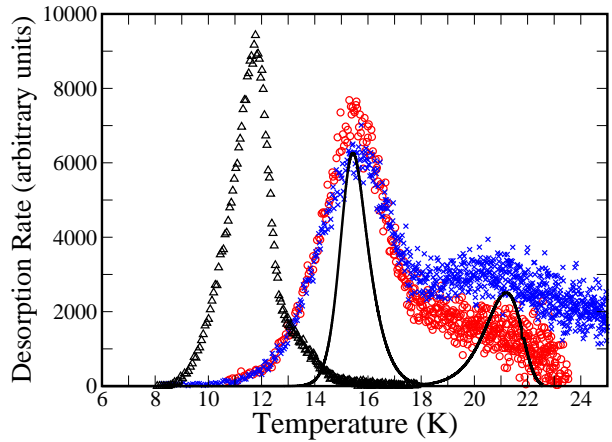}
\caption{
TPD curves of HD desorption
after irradiation of 
HD molecules ($\times$) 
and 
H+D atoms ($\circ$)
on amorphous silicate.
The H+D results are fitted using the rate equation model
(solid line). 
Also shown, for comparison, 
is HD desorption after irradiation
with H+D atoms on polycrystalline silicate ($\triangle$). 
}
\label{fig:1}
\end{center}
\end{figure}

The results were analyzed using rate equation models.
The energy barriers for the diffusion and desorption
processes were obtained
(Katz et al. 1999, Cazaux \& Tielens 2004, Perets et al. 2005).
Using these parameters,
the conditions for efficient H$_{2}$
formation on different astrophysically relevant surfaces
were found.
In particular, the
formation of H$_{2}$ on polycrystalline silicates
was found to be efficient only in
a narrow temperature window below 10K.
Since the typical dust grain temperature in diffuse interstellar
clouds is higher than 10K,
these results
indicated that polycrystalline silicate grains cannot be efficient
catalysts for H$_{2}$ formation in most diffuse clouds.

Here we present
experiments on molecular hydrogen formation 
on \emph{amorphous} silicates 
and analyze the results using a suitable rate equation model
(\cite{Perets2005}). 
Using the parameters that best fit the experimental
results, the efficiency of hydrogen recombination on grains
is obtained for 
a range of conditions pertinent to diffuse interstellar clouds. 
It is found that
unlike the polycrystalline silicate grains,
amorphous silicate grains, which are the main silicate
component in interstellar clouds,
are efficient catalysts for H$_{2}$ formation within a broad temperature
window that extends at least up to about 14K
(Perets et al. 2007).

In the experiments reported here, we 
used beams of low
fluxes and short dosing times.
Using the standard Langmuir-Hinshelwood analysis,
plotting the total yield of HD vs. the exposure time
(\cite{Perets2007})
we estimated the coverage to be a small fraction 
(a few percent) of a monolayer (ML).
This is still far from interstellar values but is within the 
regime in which results can be safely extrapolated to diffuse 
cloud conditions
(\cite{Katz1999,Perets2005}). 
The interstellar dust analogues we used are amorphous silicate
samples, 
(Fe$_{0.5}$, Mg$_{0.5}$)$_{2}$SiO$_{4}$, 
prepared by laser ablation
(\cite{Brucato2002}). 

\begin{figure}[ht]
\begin{center}
\vspace{0.2in}
\includegraphics[width=8 cm]{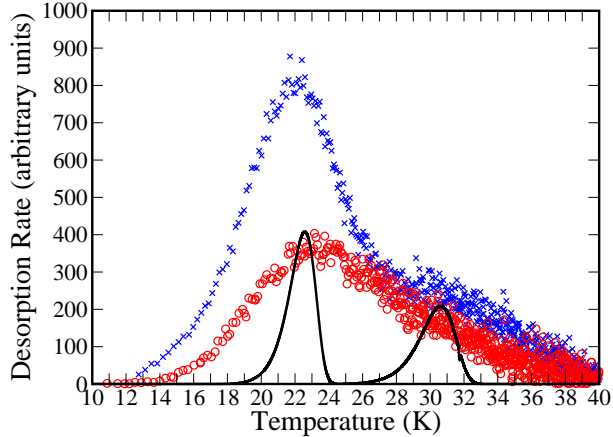}
\caption{
TPD curves of HD desorption after irradiation of
H+D atoms ($\circ$) and HD molecules ($\times$)
on the amorphous silicate sample at surface
temperature of 10K. 
The H+D data was fitted using the rate equation model
(solid line).
\label{fig:2}
}
\end{center}
\end{figure}

The experiment consists of adsorbing hydrogen atoms onto the surface
while monitoring the amount of hydrogen molecules that are
formed. 
To increase the signal to noise ratio, hydrogen and deuterium
atoms are used and the formation of HD is monitored. 
The measurement of HD formation is done in two steps. 
First, we record the amount of HD that forms and comes off 
the surface while the sample is being dosed with H and D atoms 
(the \emph{irradiation phase}). 
Next, after dosing is completed, in a  
TPD experiment, the surface temperature
is raised rapidly and the rate of HD desorption
is measured
(the \emph{TPD phase}). 
By far, the main contribution comes from the TPD phase.

Irradiations with beams of H and D 
(\char`\"{}H+D\char`\"{} thereafter) 
were done on an amorphous silicate surface, 
at a surface temperature of $T_{0} \simeq 5.6$K, 
and the formation of HD molecules was measured.
In a separate experiment, a beam
of HD molecules was irradiated on the same surface.
During the TPD runs, the sample temperature was monitored as a function of
time. 
The desorption rates of HD molecules vs. surface temperature during the
TPD runs are shown in Fig. 1,  
for H+D irradiation on polycrystalline silicate ($\triangle$)
and amorphous silicate ($\circ$)
surfaces, with irradiation times of 120 s.
The TPD curve following 
irradiation of HD molecules on an amorphous silicate 
surface
is also shown 
($\times$). 
The results of current experiments of H+D irradiation on amorphous silicates,
clearly differ from those of earlier experiments
on polycrystalline silicates.
The desorption curves from amorphous silicates contain two wide peaks,
located at a significantly higher temperatures than the single narrow
peak obtained for the polycrystalline silicate.
The higher peak temperatures 
indicate that the relevant energy
barriers are larger, while their large width
reflects a broader distribution 
of the energy barriers of the HD desorption
sites. 
The TPD curve of HD desorption from amorphous silicates, after irradiation
with HD molecules (crosses in Fig. 1),
is qualitatively similar to the curve obtained for H+D irradiation.
In particular, the peak temperatures are the same.
The relative weights of the high temperature peaks vs. the
low temperature peaks are somewhat different.
Also, in similar experiments with higher values of $T_0$
(Vidali et al. 2007),
a third peak was observed at higher temperatures
(Fig. 2).
We attribute this behavior to diffusion of HD molecules, 
which gradually migrate from shallow into
deep adsorption sites 
(\cite{Perets2005,Amiaud2006}).
An analysis of experimental data for irradiation at 
higher temperature and from amorphous silicates with 
other Mg/Fe ratios wiil be given elsewhere.

The experimental results were fitted using
the rate equation model described in 
Perets et al. (2005).
The fits are shown as
solid lines in Figs. 1 and 2. 
When a Gaussian distribution of activation energies
is introduced around the reported barrier for each process, 
the solid lines follow the experimental traces much more 
closely 
(Perets et al. 2007).
The parameters for the diffusion
and desorption of hydrogen atoms and molecules 
on the amorphous silicate surface were obtained. 
These include the energy barrier 
$E_{\rm H}^{\rm diff}$
for the diffusion of H atoms
and the barrier
$E_{\rm H}^{\rm des}$  
for their desorption.
The value obtained for the desorption barrier
of H atoms
should be considered only as a lower bound, 
because the TPD results are insensitive 
to variations in 
$E_{\rm H}^{\rm des}$,  
as long as it is higher than the reported value.
A justification for not considering explicitly 
the two isotopes (H and D) is given in 
Vidali et al. (2007). 
The desorption barriers of HD molecules adsorbed
in shallow (lower temperature peak)
and deep (higher temperature peaks) 
sites, are given by 
$E_{\rm HD}^{\rm des}(j),$ 
where $j=1$, 2 and 3, respectively.
The density of adsorption sites,
$s=7 \times 10^{14}$ (cm$^{-2}$) 
on the amorphous silicate sample,
was also obtained
(\cite{Perets2007}).

The recombination efficiency is defined
as the fraction of hydrogen atoms adsorbed on the surface which come
out as molecules. 
Using the parameters obtained from the experiments, 
we evaluated
the recombination efficiency on 
amorphous silicate surfaces
under interstellar conditions
as a function of the grain temperature. 
The incoming flux was taken as
5.2 $\times$ $10^{-10}$ (ML s$^{-1}$). 
This flux
corresponds to gas density of 10 
(atoms cm$^{-3}$) and 
gas temperature of 100K.  
\begin{table}
\caption{
Parameters for molecular hydrogen formation:
the diffusion and desorption barriers
of H atoms, the desorption barrier of HD molecules
and the density of adsorption sites.
}
\vspace{0.1in}
\begin{centering}
\begin{tabular}{lcccc}
\hline 
Material &
$E_{\rm H}^{\rm diff}$(meV) &
$E_{\rm H}^{\rm des}$(meV)  &
$E_{\rm HD}^{\rm des}$(meV) &
$s$ (cm$^{-2}$) \\
\hline 
Polycrystalline Silicate&
25 &
32 &
27 &
2 $\times 10^{14}$ \\
Amorphous Silicate&
35     &
44     &
35, 53, 75 & 
7 $\times 10^{14}$ \\
\hline
\end{tabular}
\par
\end{centering}
\label{t:E_barriers} 
\end{table}
A window of high recombination efficiency is found between 8-13K,
compared to 6-10K for polycrystalline silicate under similar conditions. 
For gas density of 100 
(atoms cm$^{-2}$),
the high efficiency window for the amorphous silicates 
surface shifts to 9-14K.
At higher temperatures
atoms desorb from the surface before they have sufficient time to
encounter each other. 
At lower temperatures diffusion is suppressed and
the Langmuir-Hinshelwood mechanism is no longer efficient.
Saturation of the surface with immobile
H atoms 
might render the Eley-Rideal mechanism more efficient
in producing some 
recombination 
(\cite{Katz1999,Perets2005}). 
Our results
thus indicate that the
recombination efficiency of hydrogen on 
amorphous silicates is high
in this temperature range, which is relevant to 
diffuse interstellar clouds.
Therefore, amorphous silicates 
seem to be good candidates for interstellar
grain components on which hydrogen recombines 
with high efficiency.
These results are in agreement with theoretical
predictions 
on the effects of surface roughness
(\cite{Cuppen2005}).
They also indicate
that on amorphous surfaces, 
newly formed H$_{2}$ molecules are thermalized on the surface
and do not promptly desorb. 
Consequently, H$_{2}$ molecules formed
on and desorbed from realistic 
\emph{amorphous} interstellar dust
are expected to 
have low kinetic energy and 
would probably not occupy excited
vibrational or rotational states.

This work was supported 
by NASA grants NAG5-11438 and NAG5-9093 (G.V), 
NSF grant AST-0507405 (G.V),
the Adler
Foundation for Space Research and the Israel Science Foundation (O.B)
and the
Italian Ministry for University and Scientific Research (V.P).


\begin{thebibliography}{}

\bibitem[Amiaud et al. 2006]{Amiaud2006} 
Amiaud, L., Fillion, J.H., Baouche, S., Dulieu, F., 
Momeni, A., \& Lemaire, J.L.
2006, J. Chem. Phys., 124, 094702

\bibitem[Brucato et al. 2002]{Brucato2002} 
Brucato, J.R., Mennella, V., Colangeli, L., Rotundi, A., \& Palumbo, P.
2002, Plan. Space Sci. 50, 829

\bibitem[Cazaux \& Tielens 2004]{Cazaux2004} 
Cazaux, S., \& Tielens, A.G.G.M. 
2004, ApJ, 604, 222

\bibitem[Creighan et al. 2006]{Creighan2006}
Creighan, S.C., Perry, J.S.A., \& Price, S.D.
2006, J. Chem. Phys., 124, 114701

\bibitem[Cuppen \& Herbst 2005]{Cuppen2005}
Cuppen, H. M. \& Herbst, E., 
2005, MNRAS, 361, 565

\bibitem[Dulieu et al. 2005]{Dulieu2005}
Dulieu, F., Amiaud, L., Baouche, S., Momeni, A., 
Fillion, J.H., \& Lemaire, J.L.
2005, Chem. Phys. Lett., 404, 187

\bibitem[Gould \& Salpeter 1963]{Gould1963} 
Gould, R.J., \& Salpeter, E.E. 
1963, ApJ, 138, 393

\bibitem[Hollenbach \& Salpeter 1971]{Hollenbach1971a} 
Hollenbach, D.J., \& Salpeter, E.E.	
1971, ApJ, 163, 155

\bibitem[Hollenbach et al. 1971]{Hollenbach1971b} 
Hollenbach, D.J., Werner, M.W. \& Salpeter, E.E.	
1971, ApJ, 163, 165

\bibitem[Hornekaer et al. 2003]{Hornekaer2003} 
Hornekaer, L., Baurichter, A., Petrunin, V.V., 
Field, D., \& Luntz, A.C.
2003, Science, 302, 1943

\bibitem[Hornekaer et al. 2005]{Hornekaer2005} 
Hornekaer, L., Baurichter, A., Petrunin, V.V., 
Luntz, A.C., Kay, B.D., \& Al-Halabi, A.
2005, J. Chem. Phys., 122, 124701

\bibitem[Katz et al. 1999]{Katz1999} 
Katz, N., Furman, I., Biham, O., Pirronello, V., \& Vidali, G. 
1999, ApJ, 522, 305

\bibitem[Manico et al. 2001]{Manico2001} 
Manic\'o, G., Ragun\'i, G., Pirronello, V., Roser, J.E., \& Vidali, G.
2001, ApJ, 548, L253

\bibitem[Perets et al. 2005]{Perets2005}
Perets, H.B., Biham, O., Manic\'o, G., Pirronello, V., 
Roser, J., Swords, S., \& Vidali, G.
2005, ApJ, 627, 850

\bibitem[Perets et al. 2007]{Perets2007}
Perets, H.B., Lederhendler, A., Biham, O., Vidali, G., 
Li, L., Swords, S., Congiu, E., Roser, J., Manic\'o, G., 
Brucato, J.R., \& Pirronello, V.
2007, ApJ, 661, L163

\bibitem[Pirronello et al. 1997]{Pirronello1997} 
Pirronello, V., Biham, O., Liu, C., Shen, L., \& Vidali, G.
1997, ApJ, 483, L131

\bibitem[Pirronello et al. 1999]{Pirronello1999} 
Pirronello, V., Liu, C., Roser, J.E., Vidali, G.
1999, A\&A, 344, 681

\bibitem[Roser et al. 2002]{Roser2002} 
Roser, J.E., Manic\'o, G., Pirronello, V., \& Vidali, G.
2002, ApJ, 581, 276

\bibitem[Vidali et al. 2007]{Vidali2007}
Vidali, G., Pirronello, V., Li, L.,
Roser, J., Manic\'o, G.,
Mehl, H., Lederhendler, A.,
Perets, H.B., Brucato, J.R., \& Biham, O.
2007, J. Phys. Chem., submitted

\bibitem[Williams 1968]{Williams1968} 
Williams D.A. 
1968, ApJ, 151, 935

\bibitem[Williams et al. 2007]{Williams2007} 
Williams, D.A., Brown, W.A., Price, S.D., Rawlings, J.M.C., \& Viti, S.
2007, Astron. Geophys., 48, 25

\end{thebibliography}
\end{document}